# AN IMPROVED VARIABLE STEP-SIZE ZERO-POINT ATTRACTING PROJECTION ALGORITHM

*Jianming Liu and Steven L Grant*[1]

Department of Electrical and Computer Engineering, Missouri University of Science and Technology, Rolla, Missouri 65409.

## ABSTRACT

This paper proposes an improved variable step-size (VSS) scheme for zero-point attracting projection (ZAP) algorithm. The proposed VSS is proportional to the sparseness difference between filter coefficients and the true impulse response. Meanwhile, it works for both sparse and non-sparse system identification, and simulation results demonstrate that the proposed algorithm could provide both faster convergence rate and better tracking ability than previous ones.

*Index Terms*—Variable step-size, zero-point attracting projection, adaptive filter, sparse system identification

## 1. INTRODUCTION

In the sparse system identification problem, such as the network echo cancellation, only a small percentage of coefficients are active and most of the others are zero or close to zero. Considering that the classical least-mean-square (LMS) algorithm is slow for sparse system identification [1], the family of proportionate algorithms has been proposed to exploit the sparse nature of the system to improve performance [2]-[4]. Besides to that, a new kind of method, zero-point attracting projection (ZAP), has been recently proposed to solve sparse system identification problem. The zero-attracting LMS (ZA-LMS) algorithm uses an $l_1$ norm penalty in the standard LMS cost function [5] and $l_0$ norm LMS was proposed in [6] too. When the solution is sparse, the gradient descent recursion will accelerate the convergence of near-zero coefficients of the sparse system.

The above scheme was referred as *zero-point attraction projection* (ZAP) in [7]. The performance analysis of ZA-LMS has been report in [8]-[10], and analysis showed that the step-size of the ZAP term denotes the importance or the intensity of attraction. A large step-size for ZAP results in a faster convergence, but the steady-state misalignment also increases. So, the step-size of ZAP is also a trade-off between convergence rate and steady-state misalignment, which is similar to the step-size trade-off of LMS.

There are some theoretical results about variable step-size ZAP but they could not be calculated in practice [9]-[11]. One practical variable step-size ZAP was proposed by You, et al. in [12], and You's VSS ZAP was simply initialized to be a large value and reduced by a factor when the algorithm is convergent. However, this heuristic strategy cannot track the change in the system response due to the very small steady-state step-size.

Another better VSS-ZAP was proposed in [13], in which a variable step-size based on the gradient of estimated filter coefficients' sparseness was proposed and the gradient is approximated by the difference between the sparseness measure of current filter coefficients and an averaged sparseness measure. This variable step-size ZAP works in the way of being an indicator whether the current filter's sparseness has reached the steady-state instead of measuring the real sparseness difference between the filter and true system response. Meanwhile, in this paper, a new variable step-size ZAP is proposed by defining the sparseness distance, then the proposed VSS is determined systematically based on sparseness difference between filter coefficients and true impulse response.

This paper is organized as follows. Section 2 reviews the recently VSS algorithms for ZAP, and in Section 3 we present the proposed VSS ZA-LMS algorithm. The simulation results and comparison to the previous VSS algorithms are presented in Section 4. Finally conclusions are drawn in Section 5.

## 2. REVIEW OF VSS ZAP

In this section, we will review the ZAP algorithm and the variable step-size ZAP algorithms in previous literature.

### 2.1. Introduction to ZAP

Consider a linear system with its input $x(n)$ and output $d(n)$ related by

$$d(n) = \mathbf{x}^T(n)\mathbf{h}(n) + v(n), \qquad (1)$$

where $\mathbf{x} = [x(n)\,x(n-1)\cdots x(n-L+1)]^T$ is the input vector, $\mathbf{h} = [h_0\,h_1\cdots h_{L-1}]^T$ is unknown system with length $L$, and $v(n)$ is the additive noise which is independent with $\mathbf{x}(n)$. The estimation error of the adaptive filter output with respect to the desired signal is defined as

$$e(n) = d(n) - \mathbf{x}^T(n)\mathbf{w}(n-1). \qquad (2)$$

This error, $e(n)$ is used to adapt the adaptive filter $\mathbf{w}(n)$. The ZA-LMS algorithm with $l_1$ norm constraint was proposed in [6], and its update equation is

$$\mathbf{w}(n) = \mathbf{w}(n-1) + \mu\mathbf{x}(n)e(n) - \kappa\,\mathrm{sgn}(\mathbf{w}(n-1)), \qquad (3)$$

in which $\mu$ is the step-size of adaption, $\kappa$ is the step-size of zero attractor, and $\mathrm{sgn}(\cdot)$ is a component-wise sign function defined as

$$\mathrm{sgn}(x) = \begin{cases} \dfrac{x}{|x|}, & x \neq 0; \\ 0, & \text{elsewhere}. \end{cases} \qquad (4)$$

### 2.2. Review of Variable Step-size ZAP Algorithms

The variable step-size for ZAP used in [12] is rather direct: $\kappa$ is initialized to be a large value, and reduced by a factor $\eta$ when the algorithm is convergent. This reduction is conducted until is sufficiently small, i.e. $\kappa < \kappa_{\min}$, which means that the error reaches a low level. However, as mentioned in the introduction, this heuristic strategy will not react to a change in the system response since it will get stuck due to the very small steady-state step-size.

Therefore, in order to solve this issue, a new variable step-size ZAP algorithm was proposed in [13] by us, which is based on the measurement of the sparseness gradient approximated by the difference between the sparseness measure of current filter coefficients and an averaged sparseness measurement as below.

The averaged sparseness measure could be estimated adaptively with a forgetting factor $\lambda$:

$$\phi(n) = (1-\lambda)\phi(n-1) + \lambda J(\mathbf{w}(n)),\ 0 < \lambda < 1, \qquad (5)$$

where $J(\mathbf{w}(n))$ is a sparseness measure of the filter coefficients, and we will use the following $l_1$ norm sparseness measure through this paper

$$J(\mathbf{w}(n)) = \|\mathbf{w}(n)\|_1 = \sum_{i=1}^{L}|w_i(n)|. \qquad (6)$$

The difference between the sparseness measure of current filter coefficients and the averaged sparseness measurement is calculated by:

$$\delta(n) = J(\mathbf{w}(n)) - \phi(n-1) \qquad (7)$$

In order to obtain a good and stable estimate of the gradient, a long-term average using infinite impulse response filters is used to calculate the proposed variable step-size

$$\kappa(n) = (1-\alpha)\kappa(n-1) + \alpha\gamma\delta(n),\ 0 < \alpha < 1. \qquad (16)$$

As mentioned in the introduction, this variable step-size ZAP indicates whether the current filter's sparseness has reached the steady-state instead measuring the sparseness distance between the filter and real system. Therefore, we will propose a variable step-size algorithm for ZA-LMS which is derived based on the difference between current filter coefficients' sparseness and the real sparseness in next section.

### 3. PROPOSED VSS ZA-LMS

In this section, we will propose the variable step-size ZAP, and further improve its performance for non-sparse system identification.

### 3.1. The Proposed Scheme of Variable Step-size ZAP

Our proposed new variable step-size ZAP algorithm is based on the idea that the step-size should be proportional to the sparseness distance which is defined as the difference between the sparseness measure of current filter coefficients and real sparseness of the system. Based on $l_1$ norm, we define the following averaged sparseness distance

$$\delta(n) = \frac{1}{L}\left|\|\mathbf{w}(n)\|_1 - \|\mathbf{h}(n)\|_1\right| = \frac{1}{L}\left|\sum_{i=1}^{L}|w_i(n)| - \sum_{i=1}^{L}|h_i(n)|\right|. \qquad (8)$$

Then we rewrite (8) as

$$\delta(n) = \frac{1}{L}\left|\mathbf{h}^T(n)\mathrm{sgn}(\mathbf{h}(n)) - \mathbf{w}^T(n)\mathrm{sgn}(\mathbf{w}(n))\right|. \qquad (9)$$

However, considering the real system is unknown, we argue that $\mathrm{sgn}(\mathbf{h}(n))$ could be approximated by $\mathrm{sgn}(\mathbf{w}(n))$. This assumption is acceptable because it holds for the coefficients with large magnitude, and for the small and unstable coefficients close to zero, considering that its magnitude is relatively small, it will not cause large error in the approximation. We will verify the performance of this assumption in the simulation section later, and using this assumption in (9), we have

$$\delta(n) \approx \frac{1}{L}\left|(\boldsymbol{h}(n)-\boldsymbol{w}(n))^T \operatorname{sgn}(\boldsymbol{w}(n))\right|$$
$$= \frac{1}{L}\left|\Delta \boldsymbol{h}^T(n)\operatorname{sgn}(\boldsymbol{w}(n))\right|. \quad (10)$$

The system mismatch is defined as $\Delta \boldsymbol{h}(n) = \boldsymbol{h}(n) - \boldsymbol{w}(n)$. Using the similar approximation in [14], we have

$$\left|\Delta \boldsymbol{h}^T(n)\operatorname{sgn}(\boldsymbol{w}(n))\right|$$
$$\approx L \left|\frac{\Delta \boldsymbol{h}^T(n)\boldsymbol{x}(n)\boldsymbol{x}^T(n)\operatorname{sgn}(\boldsymbol{w}(n))}{\boldsymbol{x}^T(n)\boldsymbol{x}(n)}\right|. \quad (11)$$

It should be noted that we use the following assumptions in [14]

$$\boldsymbol{R}_{xx}(n) = \boldsymbol{x}(n)\boldsymbol{x}^T(n) \approx \sigma_x^2 \boldsymbol{I}, \text{ and } \boldsymbol{x}^T(n)\boldsymbol{x}(n) \approx L\sigma_x^2. \quad (12)$$

Furthermore, the residual error is defined as

$$\varepsilon(n) = \Delta \boldsymbol{h}^T(n)\boldsymbol{x}(n). \quad (13)$$

Substituting (11) and (13) into (10), we could rewrite (10) as

$$\delta(n) \approx \left|\frac{\varepsilon(n)\boldsymbol{x}^T(n)\operatorname{sgn}(\boldsymbol{w}(n))}{\boldsymbol{x}^T(n)\boldsymbol{x}(n)}\right|. \quad (14)$$

However, the residual error in (14) is still unknown, but similar to [13], to avoid over-shoot, a long-term time average should be used to calculate the proposed variable step-size as below

$$\kappa(n) = (1-\alpha)\kappa(n-1) + \alpha\gamma\delta(n), \ 0 < \alpha < 1, \quad (15)$$

in which $\alpha$ is a smoothing factor and $\gamma$ is a correction factor. Meanwhile, considering the additive noise is independent with input, the cross-correlation between the input and residual error is the same as the cross-correlation between input and error. Therefore, we could replace the residual error in (14) with the error signal, which gives us

$$\delta(n) \approx \left|\frac{e(n)\boldsymbol{x}^T(n)\operatorname{sgn}(\boldsymbol{w}(n-1))}{\boldsymbol{x}^T(n)\boldsymbol{x}(n)}\right|. \quad (16)$$

### 3.2. Improved Variable Step-size ZAP for Both Sparse and Non-sparse System

Besides to the $l_1$ norm sparseness measures defined in (6), another popular measurement of channel sparsity was used in [13], and for a channel $\boldsymbol{h}(n)$, its sparsity $\xi(\boldsymbol{h}(n))$ can be defined as

$$\xi(\boldsymbol{h}(n)) = \frac{L}{L - \sqrt{L}}\left(1 - \frac{\|\boldsymbol{h}(n)\|_1}{\sqrt{L}\|\boldsymbol{h}(n)\|_2}\right), \quad (17)$$

where $L > 1$ is the length of the channel $\boldsymbol{h}(n)$, and $\|\boldsymbol{h}(n)\|_1$ and $\|\boldsymbol{h}(n)\|_2$ are the $l_1$ norm and $l_2$ norm of $\boldsymbol{h}(n)$. The value of $\xi(\boldsymbol{h}(n))$ is between 0 and 1. For a sparse channel the value of sparsity is close to 1 and for a dispersive channel, this value is close to 0. In [13], this property was used to remove the ZAP term when the channel is dispersive, which is preferable.

We could also take advantage of this property and propose the following averaged sparseness distance as variable step-size for ZA-LMS

$$\delta(n) = \frac{1}{L}\left|\xi(\boldsymbol{h}(n)) - \xi(\boldsymbol{w}(n))\right|$$
$$= \frac{1}{L(\sqrt{L}-1)}\left|\frac{\|\boldsymbol{h}(n)\|_1}{\|\boldsymbol{h}(n)\|_2} - \frac{\|\boldsymbol{w}(n)\|_1}{\|\boldsymbol{w}(n)\|_2}\right|. \quad (18)$$

We assume the gain of the real channel and filter coefficients are the same, i.e.

$$\|\boldsymbol{h}(n)\|_2 \approx \|\boldsymbol{w}(n)\|_2. \quad (19)$$

However, this assumption might not be accurate, especially at the initial phase of the adaption. Therefore, a reasonable minimum threshold of $\|\boldsymbol{w}(n)\|_2$ should be used to avoid this issue. Then we could further simplify (19) as

$$\delta(n) \approx \frac{1}{L(\sqrt{L}-1)}\frac{1}{\|\boldsymbol{w}(n)\|_2}\left|\|\boldsymbol{h}(n)\|_1 - \|\boldsymbol{w}(n)\|_1\right|. \quad (20)$$

Considering (16), we obtain the proposed variable step-size for ZA-LMS which could work for both dispersive and sparse channel as below

$$\delta(n) \approx \frac{1}{\sqrt{L}-1}\frac{1}{\|\boldsymbol{w}(n)\|_2}\left|\frac{e(n)\boldsymbol{x}^T(n)\operatorname{sgn}(\boldsymbol{w}(n))}{\boldsymbol{x}^T(n)\boldsymbol{x}(n)}\right|. \quad (21)$$

## 4. SIMULATION RESULTS

In this section, we do the results of computer simulations in the scenario of echo cancellation. We use both sparse impulse response and a dispersive random impulse response. They are both with the same length, *L=512*, and the LMS adaptive filter is with the same length.

The convergence state of adaptive filter is evaluated using the normalized misalignment which is defined as

$$20\log_{10}(\|\boldsymbol{h}-\boldsymbol{w}\|_2 / \|\boldsymbol{h}\|_2) \quad (22)$$

The input is white Gaussian noise signal and independent white Gaussian noise is added to the system background with a signal-to-noise ratio, SNR = 30 dB.

In the first simulation, we would like to verify the performance of the approximation $\text{sgn}(\boldsymbol{h}(n)) = \text{sgn}(\boldsymbol{w}(n))$ in (10) as in Fig. 1. In order to demonstrate the tracking ability, there is an echo path change at sample 5000 by switching from one sparse impulse response to another sparse impulse response. It is observed that, even though the approximation is not very accurate in the initial phase, it could be very good for tracking the change of the echo path. This is predictable since the filter coefficients are initialized as zeros, then there will be larger difference between $\text{sgn}(\boldsymbol{h}(n))$ and $\text{sgn}(\boldsymbol{w}(n))$. However, this assumption is still good enough for the application scenario of proposed variable step-size ZAP, which will be verified by the following simulations.

In the second simulation, we compare the proposed VSS algorithm to LMS, fixed step-size ZA-LMS, You's VSS in [12] and Liu's VSS in [13] for sparse system identification. It should be noted that sparseness measure (17) is used in Liu's VSS, and (21) is used as the proposed variable step-size. Meanwhile, to evaluate the performance of the tracking ability, there is also an echo path change at sample 5000, and according to the simulation result in Fig. 2, the parameters of the variable step-size are intentionally set to have similar steady-state misalignment for the first adaption before echo path change. It is observed that, because You's VSS cannot react to echo path change, it could only obtain similar tracking performance with original ZAP. Meanwhile, Liu's VSS and proposed VSS could track the echo path change quickly, and the proposed VSS outperforms the previous ones.

Next, in order to demonstrate the performance for dispersive channel, we switch one dispersive impulse response to another dispersive response at sample 5000, and use the same VSS algorithms and parameters as the second simulation. As shown in Fig. 3, it is clear that the proposed VSS ZAP could also obtain much better tracking performance under non-sparse system than previous ones and avoid the possible performance degradation.

## 5. CONCLUSION

An improved variable step-size zero-point attraction projection algorithm was proposed based on the estimation of $l_1$ sparseness distance, which could work for both sparse and non-sparse system identification. Simulation results verify that the proposed VSS ZAP could provide better tracking ability than previous VSS ZAP algorithms for both sparse and non-sparse system identification.

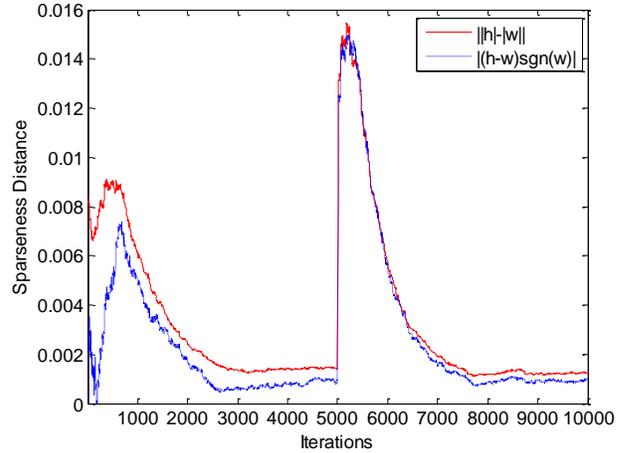

**Fig.1** Performance demonstration of approximation $\text{sgn}(\boldsymbol{h}(n)) = \text{sgn}(\boldsymbol{w}(n))$ in (10).

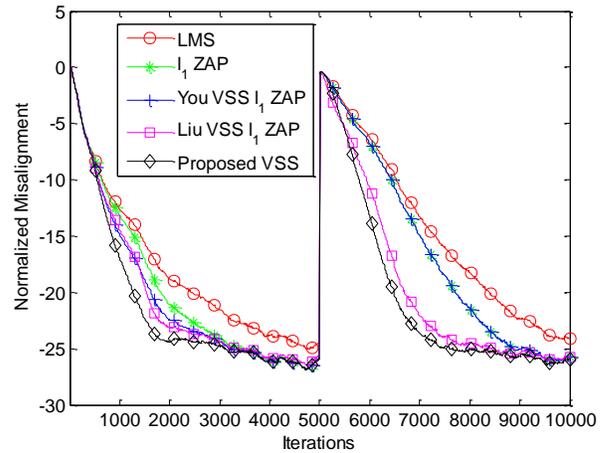

**Fig.2** Comparison of normalized misalignment for sparse system identification.

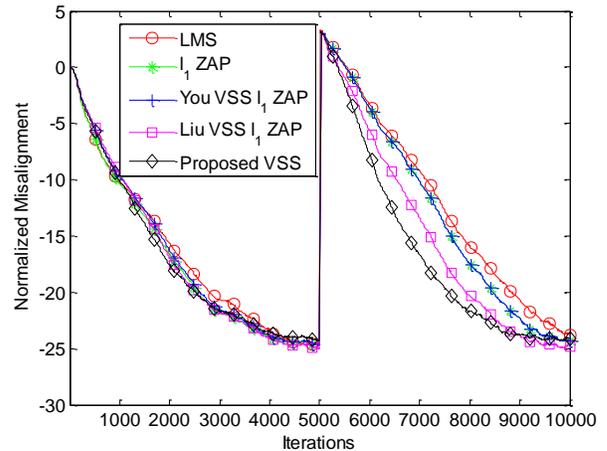

**Fig.3** Comparison of normalized misalignment for dispersive system identification.